\documentstyle[11pt,sf2000]{article}

\begin{document}

\title{Probing Planet Formation Among Nearby Young Stars}
\author{Ray Jayawardhana}
\affil{University of California, Berkeley}

\section{Introduction}
The recent identification of several groups of young stars within 100 parsecs
of the Sun has generated widespread interest. Given their proximity and
possible age differences, these systems are ideally suited for detailed
studies of disk evolution and planet formation. Here I discuss recent results
and prospects for the near-future.

\section{TW Hydrae Association}
At a distance of $\sim$ 55 pc, the TW Hydrae Association (Kastner et al. 1997) 
is almost three times closer than the Taurus molecular 
clouds, and its estimated age of $\sim$10 Myr fills a significant gap in the 
age sequence between previously
known 1-Myr-old T Tauri stars and 50-Myr-old nearby open clusters. The group
consists mostly of low-mass stars, typically a few tenths as massive as our
sun, and includes several binary systems as well as one remarkable quadruple
system (HD 98800). There is only one early-type star (HR 4796A).
The origin of the
TW Hydrae Association remains a bit of a mystery. There is no obvious parent
cloud and the stars are dispersed across some 20 degrees on the sky and 20 pc
in radial distance making it difficult to determine their birthplace.
Nevertheless, they do offer the prospect of better constraining timescales for
disk evolution and planetary birth.

Over the past two years, we have obtained mid-infrared observations of 
TW Hya stars using the OSCIR instrument on the Keck II and CTIO 4-meter 
telescopes.  During this program, we imaged a spatially-resolved dust disk 
around the young A star HR 4796A (Jayawardhana et al. 1998). The surface 
brightness distribution of the disk is consistent with the presence of an 
inner disk hole of $\sim$50 AU radius, as was first suggested by Jura et al. 
(1993) based on the infrared spectrum. Our results for HR 4796A and other 
stars in the group indicate rapid evolution of inner disks 
(Jayawardhana et al. 1999b). However, it is unlikely 
that there is a universal evolutionary timescale for protoplanetary 
disks, especially when the influence of companion stars is taken into account.
For example, we have detected thermal emission from a dusty disk around the 
primary, but not the secondary, in the Hen 3-600 binary system (Jayawardhana
et al. 1999a). Comparison with
the median spectral energy distribution of classical T Tauri stars suggests
that the disk around Hen 3-600A may be truncated by the secondary.

\section{Other Nearby Groups}
The all-sky survey done by the ROSAT satellite has been particularly 
useful in identifying isolated young stars through their X-ray emission. 
Of the other recently discovered stellar groups, MBM12 and Eta
Chamaeleontis appear particularly interesting. At about 65 pc, MBM12 is the
nearest known star-forming cloud, containing only 30-100 solar masses of gas.
It does not appear to be gravitationally bound, and may be breaking up on a
timescale comparable to the sound-crossing time. Thus, in a few million
years, the young stars in MBM12 may appear to be isolated objects, not
associated with any cloud material --very similar to how the TW Hydrae
stars appear at present. Hearty et al. (2000) have identified eight
low-mass young stars associated with MBM12. Most of them are classical T
Tauri stars, and are likely to be a younger population than the TW Hydrae
members. Eta Cha is a cluster of a dozen young stars, first identified in
X-rays (Mamajek et al. 1999). As with the TW Hydrae group, Eta Cha is far
from any substantial cloud. However, its members are much less dispersed than
the TW Hydrae stars, and may represent an epoch intermediate between MBM12 and
TW Hydrae Association. The comparison of disk statistics and properties
among MBM12, $\eta$ Cha, and the TW Hya Association may provide additional
insight on disk evolution.

\section{Prospects for Imaging Young Planets}
If planets have indeed formed around these stars, it may be possible to
detect them using large ground-based telescopes. Adaptive optics
allows one to search within
several AU of these nearby stars for planets a few times as massive
as Jupiter (Jayawardhana 2000). Since newborn planets are quite warm, 
such objects would be
sufficiently luminous to be detected at distances of 50-100 pc.
In other words, we should be able to look for newborn giant planets
located at distances from their parent stars similar to those of giant planets
in our own solar system. Already, at least one brown dwarf has been found in
the TW Hydrae Association (Lowrance et al. 1999) and searches for still
lower-mass objects are underway (e.g., Neuhauser et al. 2000).

\end{document}